\documentclass[conference]{IEEEtran}

\IEEEoverridecommandlockouts

\usepackage[letterpaper, top=0.75in, bottom=1.05in, left=0.625in, right=0.625in]{geometry}

\usepackage{cite}
\usepackage{amsmath,amssymb,amsfonts}
\usepackage{algorithmic}
\usepackage{graphicx}
\usepackage{textcomp}
\usepackage{xcolor}
\usepackage{enumitem}
\usepackage{algorithm,algorithmic}
\usepackage{amsthm}
\usepackage[normalem]{ulem}
\usepackage[dvipsnames]{xcolor}  

\newtheorem{propos
ition}{Proposition}

\def\BibTeX{{\rm B\kern-.05em{\sc i\kern-.025em b}\kern-.08em
 T\kern-.1667em\lower.7ex\hbox{E}\kern-.125emX}}

\IEEEaftertitletext{\vspace{-2em}}
\begin{document}

\title{Multi-Layer Secret Sharing for Cross-Layer Attack Defense in 5G Networks: a COTS UE Demonstration
\thanks{This work was supported in part by National Science Foundation (NSF) under grants CNS2451268, CNS2514415, ONR under grant N000142112472; and the NSF and Office of the Under Secretary of Defense (OUSD) – Research and Engineering, Grant ITE2515378, as part of the NSF Convergence Accelerator Track G: Securely Operating Through 5G Infrastructure Program.}
}

\author{\IEEEauthorblockN{Wai Ming Chan}
\IEEEauthorblockA{\textit{School of Electrical, Computer}\\ \textit{and Energy Engineering} \\
\textit{Arizona State University}\\
Tempe, AZ, USA \\
{wai-ming.chan@asu.edu}}
\and
\IEEEauthorblockN{R\'{e}mi Chou}
\IEEEauthorblockA{\textit{Department of Computer}\\\textit{Science and Engineering} \\
\textit{The University of Texas at Arlington}\\
Arlington, TX, USA \\
{remi.chou@uta.edu}}
\and
\IEEEauthorblockN{Taejoon Kim}
\IEEEauthorblockA{\textit{School of Electrical, Computer}\\ \textit{and Energy Engineering} \\
\textit{Arizona State University}\\
Tempe, AZ, USA \\
{taejoonkim@asu.edu}}
}

\maketitle

\begin{abstract}
This demo presents the first implementation of multi-layer secret sharing on commercial-off-the-shelf (COTS) 5G user equipment (UE), operating without infrastructure modifications or pre-shared keys. Our XOR-based approach distributes secret shares across network operators and distributed relays, ensuring perfect recovery and data confidentiality even if one network operator and one relay are simultaneously lost (e.g., under denial of service (DoS) or unanticipated attacks).
\end{abstract}


\section{Introduction}
\label{sec:intro}

Enterprise and tactical 5G deployments face coordinated cross-layer attacks, ranging from physical jamming to infrastructure compromises. As 5G facilitates mission-critical applications, resilient communications become essential in contested environments \cite{ahmad2019security}.
Conventional cryptographic defenses require secure key distribution, which becomes a critical vulnerability when those channels are compromised or unavailable in contested networks.
Moreover, VPN and encryption fail when their path is blocked or keys compromised. Our secret sharing scheme splits data across three operators without requiring keys, ensuring recovery even when one operator and one relay path fail simultaneously.

\begin{figure}[hbpt!]
\centering
\includegraphics[width=0.48\textwidth]{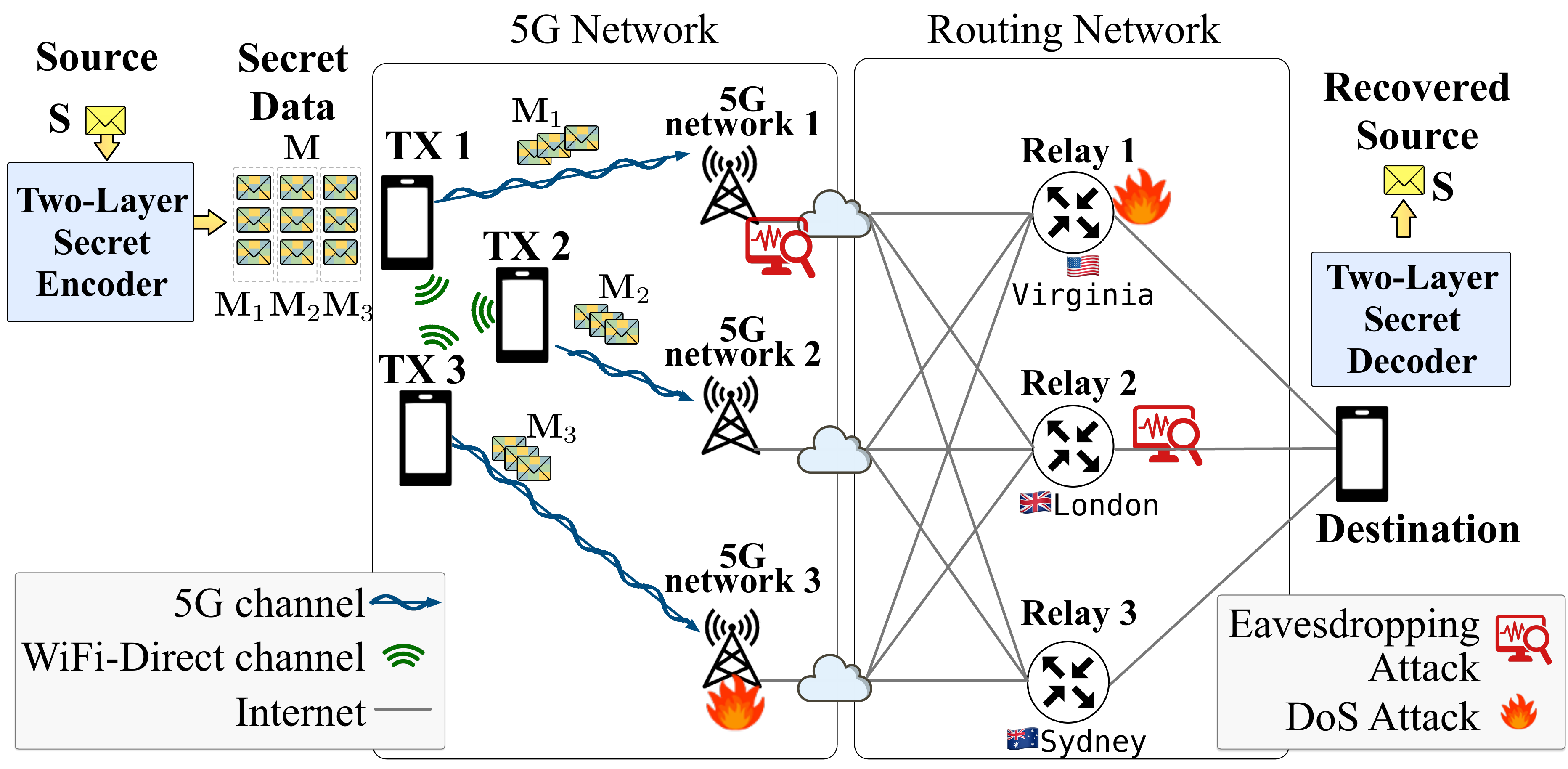}
\vspace{-20pt}  
\caption{Two-layer secret sharing system. Secret shares distributed across 3 mobile operators (columns) and 3 relays (rows) tolerate simultaneous loss of one row and one column.}
\label{fig:system}
\end{figure}

We demonstrate the first multi-layer secret sharing implementation on commercial-off-the-shelf (COTS) 5G user equipment (UE) without infrastructure modifications or pre-shared keys. As shown in Fig.~\ref{fig:system}, our system distributes secret shares across multiple mobile operators and distributed relays, enabling message recovery despite simultaneous failures at both layers. 
This demonstration validates a device-centric approach to cross-layer resilience through client-side processing.


\section{System Overview and Threat Model}
\label{sec:system}

\subsection{Threat Model}
Adversaries mount two primary attack types illustrated in Fig.~\ref{fig:system}. Eavesdropping attacks intercept secret shares at compromised infrastructure points, though individual shares reveal no information about the source message. DoS attacks disable transmission paths either by jamming a mobile operator or through compromised relays dropping packets.

\subsection{Two-Layer Encoding/Decoding Scheme}

We employ the two-layer secret sharing scheme from \cite{chan2025underreview} that encodes an $n$-bit source message $S$ into a $3\times 3$ matrix $\mathbf{M}$ of $n$-bit secret shares. The construction guarantees: (i) any $2\times 2$ submatrix suffices for full recovery, tolerating simultaneous loss of one mobile operator (column) and one relay (row), and (ii) any combination of one complete row and one complete column reveals no information about $S$. Both encoding and decoding use only XOR operations, enabling efficient implementation on resource-constrained UEs. Detailed algorithms and security proofs appear in \cite{chan2025underreview}.

\subsection{Architecture and Implementation}
A single COTS UE cannot simultaneously connect to multiple mobile operators due to eSIM switching latency and restricted dual-SIM control permissions. We overcome this limitation by coordinating three UEs via WiFi-Direct \cite{daniel2013WiFiDirect}, with each UE maintaining a dedicated connection to a different mobile operator.

\begin{figure}[hbpt!]
\centering
\includegraphics[width=0.35\textwidth]{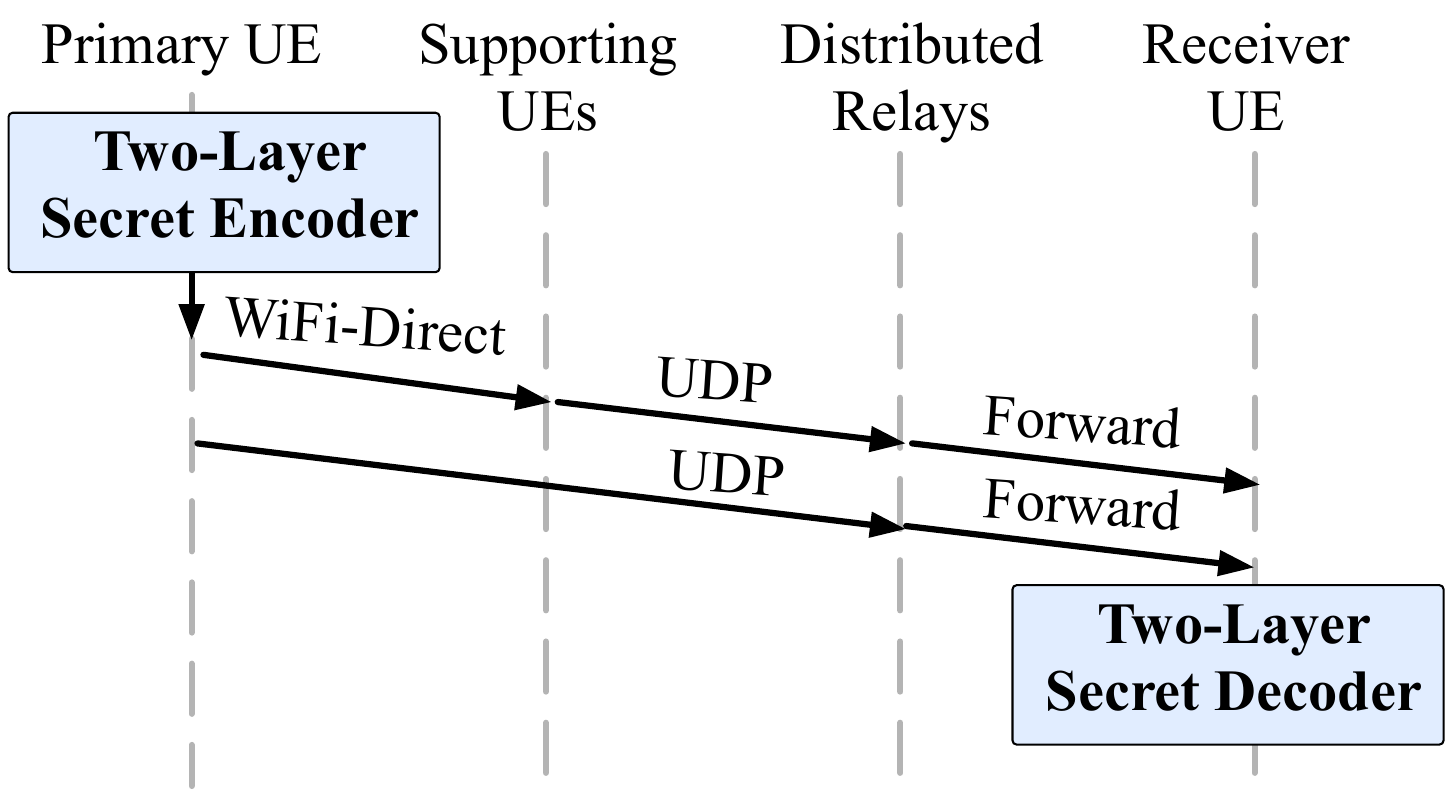}
\vspace{-10pt}  
\caption{Multi-layer secret sharing protocol. WiFi-Direct distribution followed by parallel multi-path transmission.}
\label{fig:sequence}
\end{figure}

Fig.~\ref{fig:sequence} illustrates our implementation flow. The sender encodes source message $S$ into a secret share matrix $\mathbf{M}$ and distributes columns to supporting UEs via WiFi-Direct. Each UE transmits its three secret shares through its mobile operator to designated relays, creating nine parallel paths. Secret shares are encapsulated in UDP packets with 12-byte headers (2B indices, 4B sequence, 6B message ID). Relays forward packets without inspection, and the receiver reconstructs $S$ by decoding any four secret shares forming a valid $2\times 2$ submatrix.

\section{Demonstration and Evaluation}

\subsection{Experimental Setup}

Our demonstration employs COTS 5G UEs and globally distributed relays.

\begin{figure}[!htp]
\centering
\includegraphics[width=0.34\textwidth]{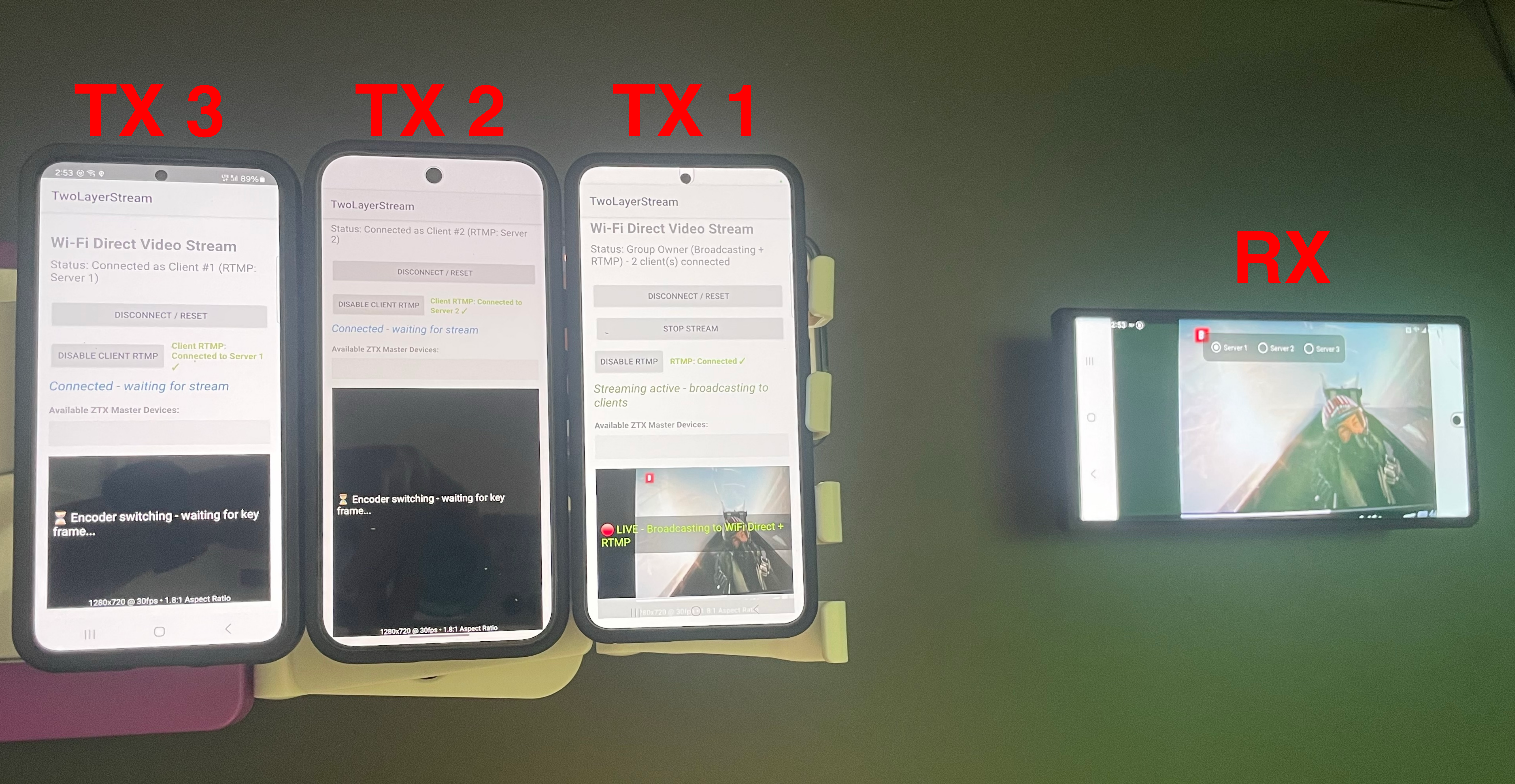}
\vspace{-10pt}  
\caption{Demo hardware setup showing coordinated sender UEs (TX1-3) and receiver (RX) with real-time decoding display.}
    \label{fig:demo_setup}
\end{figure}

\noindent\textbf{Hardware:} Fig.~\ref{fig:demo_setup} shows three sender UEs (Samsung Galaxy S25 on AT\&T and T-Mobile, Google Pixel 9 on Verizon) and one receiver UE (Galaxy S22 Ultra). Relay infrastructure comprises AWS EC2 (Virginia, U.S.), Google Cloud (London, U.K.), and Azure (Sydney, Australia) instances.

\noindent\textbf{Software:} Our Android application uses WiFi-Direct API for UE coordination and standard UDP sockets for 5G transmission. Real-time visualization displays secret share distribution and recovery status.

\vspace{-10pt} 

\vspace{-4pt} 
\subsection{Attack Simulation}
\label{subsec:attack_simulation}
Our interactive demonstration allows attendees to inject attacks and observe real-time recovery through three scenarios. \textbf{Eavesdropping attacks} mirror relay traffic to an attacker node via \texttt{tcpdump} capture. \textbf{Mobile operator DoS attack} is simulated using airplane mode to represent jamming or operator failure. \textbf{Relay DoS attack} applies \texttt{iptables} rules to drop packets at specific relays. 

\vspace{-4pt} 
\subsection{Performance Analysis}

We compare our two-layer approach to one-layer XOR secret sharing~\cite{jha2024} and simple repetition codes. The one-layer scheme~\cite{jha2024} protects against same-path failures across two layers, while repetition codes provide maximum redundancy.

\begin{figure}[htp!]
\centering
\includegraphics[width=0.35\textwidth]{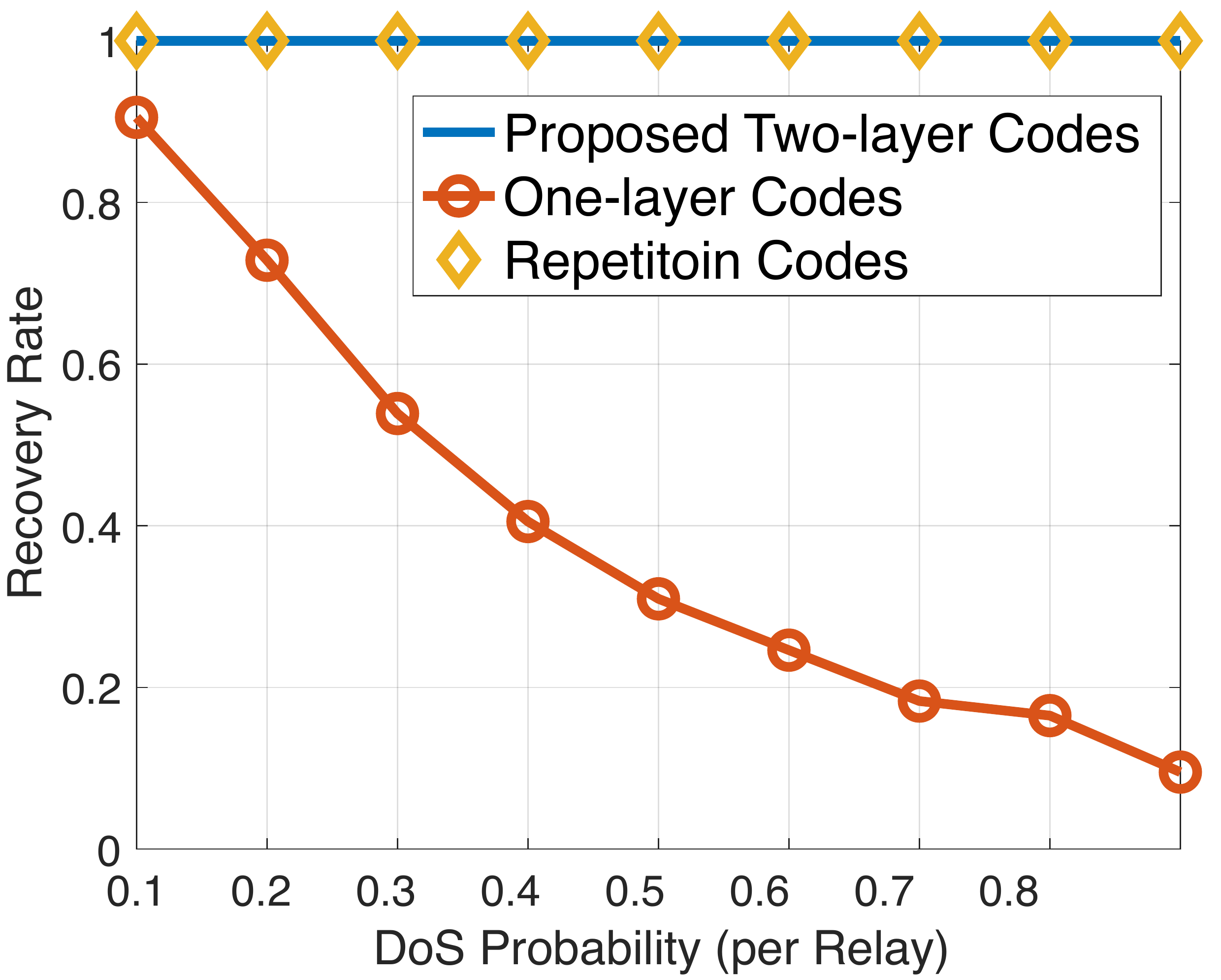}
\vspace{-10pt}  
\caption{Recovery rate under simultaneous row and column failures.}
    \label{fig:recovery_plot}
\end{figure}

Fig.~\ref{fig:recovery_plot} reveals our key advantage under the cross-layer attacks defined in Section~\ref{subsec:attack_simulation}. 
When DoS attacks target both a mobile operator and a relay simultaneously, our two-layer scheme maintains 100\% recovery. In contrast, the one-layer scheme degrades rapidly, achieving only 31\% recovery at 50\% DoS attack probability.

\vspace{-10pt}

Table~\ref{tab:tradeoff_summary} provides a comprehensive performance profile. All secret sharing schemes achieve high confidentiality ($0.9979$ entropy), but only our two-layer approach maintains perfect recovery under cross-layer attacks. This resilience incurs $60$ms additional latency compared to repetition codes, achieving both confidentiality and cross-layer protection that neither baseline method provides.

\vspace{-4pt} 
\begin{table}[htp!]
\centering
\begin{tabular}{|p{2cm}|p{2cm}|p{1.1cm}|p{1cm}|}
\hline
\textbf{Scheme} & \textbf{Confidentiality (Entropy)} & \textbf{Recovery Rate} & \textbf{Latency (ms)} \\
\hline
Two-layer Codes & \textbf{High (0.9979) }   & \textbf{100\%}& 153 \\ \hline
One-layer Codes & \textbf{High (0.9979) } & 31\% & 143 \\ \hline
Repetition Codes & {Low (0)}      & \textbf{100\%} & \textbf{93} \\ 
\hline
\end{tabular}
\vspace{0.1cm}
\caption{Performance comparison under $50\%$ DoS attacks}
\label{tab:tradeoff_summary}
\end{table}

\vspace{-12pt} 
\section{Conclusion}
This demonstration validates two-layer secret sharing as a practical defense for 5G networks using only commercial UEs. Our implementation achieves perfect message recovery and data confidentiality despite losing one mobile operator and one relay simultaneously, without requiring pre-shared keys or infrastructure modifications. While introducing some latency overhead, this approach provides cross-layer resilience unattainable through traditional cryptographic methods. Future work extends to $L$-layer implementations and explores hardware acceleration to further optimize performance.
\bibliographystyle{IEEEtran}
\bibliography{chan}

\end{document}